\documentclass[twocolumn,prl,showpacs,floatfix]{revtex4}
\usepackage{graphicx}
\begin{document}


\title{Dark Matter Caustics in Galaxy Clusters}

\author{V. Onemli$^a$ and P. Sikivie$^b$}

\affiliation{
$^a$~Department of Physics, University of Crete, 
GR-71003 Heraklion, Crete, Greece\\
$^b$~Department of Physics, University of Florida, 
Gainesville, FL 32611, USA}

\begin{abstract}
We interpret the recent gravitational lensing observations of 
Jee et al. \cite{Jee} as first evidence for a {\it caustic} ring 
of dark matter in a galaxy cluster.  A caustic ring unavoidably 
forms when a cold collisionless flow falls with net overall 
rotation in and out of a gravitational potential well.  Evidence 
for caustic rings of dark matter was previously found in the 
Milky Way and other isolated spiral galaxies.  We argue that 
galaxy clusters have at least one and possibly two or three
caustic rings.  We calculate the column density profile of a 
caustic ring in a cluster and show that it is consistent with 
the observations of Jee et al.  

\end{abstract}

\maketitle


\section{Introduction}

Using strong and weak gravitational lensing methods, Jee et al. 
\cite{Jee} constructed a column density map of the central region 
of the galaxy cluster Cl 0024+1654.  The map shows a ring of dark 
matter of radius $\simeq$ 400 kpc, width $\sim$ 150 kpc and maximum 
column density $\simeq 58 {M_\odot \over {\rm pc}^2}$.  Remarkably, 
the overdensity in dark matter is not accompanied by an analogous 
structure in x-ray emitting gas or luminous matter.  In this regard, 
Jee et al. discovered a new instance where dark and ordinary matter 
have dramatically different spatial distributions.  The well-known 
observations of the ``Bullet Cluster" 1E0657-56 \cite{bul} provided 
an earlier example.

Jee et al. interpret the dark matter ring in Cl 0024+1654 as the 
product of a near head-on collision, along the line of sight, of 
two subclusters.  They performed a simulation of the response of 
the dark matter particles to the time-varying gravitational field 
and found that, after the collision has occurred, the dark matter 
particles move outward and form shell-like structures which appear 
as a ring when projected along the collision axis \cite{Jee}.  The 
interpretation of Jee et al. fits with independent lines of evidence 
that Cl 0024+1654, an apparently relaxed cluster, is a collision of 
two subclusters.  However, it is shown in ref. \cite{ZuH} that the 
observed dark matter ring is reproduced only for highly fine-tuned, 
and hence unlikely, initial velocity distributions.

The purpose of our paper is to propose an alternative interpretation, 
to wit that Jee et al. have observed a caustic ring formed by the 
in and out flow of dark matter particles falling onto the cluster 
for the first time.  We show that the observed ring is explained 
assuming only that the dark matter falling onto the cluster has 
net overall rotation, with angular momentum vector close to the 
line of sight, and velocity dispersion less than 60 km/s.  

When cold collisionless dark matter falls from all directions into
a smooth gravitational potential well, the phase space distribution 
of the dark matter particles is characterized everywhere by a set of 
discrete flows \cite{Ips}.  The flows form outer and inner caustics.  
The outer caustics are formed by outflows where they turn around before 
falling back in.  Each outer caustic is a fold catastrophe ($A_2$) located 
on a topological sphere surrounding the potential well.  The inner caustics 
\cite{crdm,sing} are formed near where the particles with the most angular 
momentum in a given inflow reach their closest approach to the center 
before going back out.  The catastrophe structure of the inner caustics 
\cite{inner} depends on the angular momentum distribution of the infalling 
particles.  If that angular momentum distribution is characterized by 
net overall rotation, the inner caustics are rings (closed tubes) whose 
cross-section is a section of the elliptic umbilic catastrophe ($D_{-4}$) 
\cite{sing}.  These statements are valid independently of any assumptions
of symmetry, self-similarity, or anything else.

It is argued in ref.~\cite{rob} that discrete flows and caustics are 
a generic and robust property of {\it galactic} halos if the dark 
matter is collisionless and cold. The radii of the outer caustic 
spheres are predicted by the self-similar infall model \cite{FG,B} 
of halo formation.  The radii of the inner caustic rings are predicted
\cite{crdm}, in terms of a single parameter $j_{\rm max}$, after the 
model is generalized \cite{STW} to allow angular momentum for the
infalling particles.  Evidence for inner caustic rings distributed 
according to the predictions of the self-similar infall model has 
been found in the Milky Way \cite{crdm,milk,mon} and in other 
isolated spiral galaxies \cite{crdm,Kinn}.

The resolution of most present numerical simulations is inadequate 
to see discrete flows and caustics.  However such features are seen 
in dedicated simulations which increase the number of particles in the 
relevant regions of phase space \cite{Stiff,simca}.  They should also 
become apparent in fully general simulations of structure formation 
through the use of special techniques \cite{White}.

\section{Dark matter caustics in galaxy clusters}

With regard to our proposal that Jee et al. have observed a caustic 
ring of dark matter, the first question that arises is whether 
discrete flows and therefore caustics should be expected in galaxy
clusters, in view of the fact that gravitational scattering by the 
galaxies in the cluster tends to diffuse the flows.  A flow of 
particles passing through a region populated by a class of objects 
of mass $M$ and number density $n$ gets diffused by gravitational
scattering over a cone of opening angle $\Delta \theta$ whose square 
\cite{Ips}
\begin{eqnarray}
(\Delta \theta)^2 &\sim&
\int~dt~\int_{b_{\rm min}}^{b_{\rm max}}~
{4 G^2 M^2 \over b^2 v^4} n v 2 \pi b~db \nonumber\\
&\sim& 4.7 \times 10^{-3}~\left({10^3~{\rm km/s} \over v}\right)^3~
\left({M \over 10^{12} M_\odot}\right)^2\cdot\nonumber\\
&~&\cdot~\ln ({b_{\rm max} \over b_{\rm min}})~
\left({t \over 10~{\rm Gyr}}\right)~
\left({n \over {\rm Mpc}^{-3}}\right)~\ ,
\label{scat}
\end{eqnarray}
where $b$ is impact parameter, $v$ is the velocity of the flow 
and $t$ is the time over which it encountered the objects in
question.  Because the density $n$ falls off with distance $r$ 
to the cluster center as a power law ${1 \over r^\gamma}$ with 
$\gamma \geq 2$, the integral in Eq.~(\ref{scat}) is dominated 
by contributions received while passing through the central parts 
of the cluster.  To obtain an estimate, we consider a cluster 
of 200 galaxies of mass $M \sim 2 \cdot 10^{12} M_\odot$ each, 
within a region of radius 1 Mpc, and with velocity dispersion 
$\sqrt{<v^2>} \simeq 1300$ km/s.  These properties are descriptive 
of Cl 0024+1654 \cite{Czos}.  The velocity of the flow going 
through the central parts for the first time is approximately 
$v \sim 2 \sqrt{<v^2>}$.  The time to fall through the central 
region is approximately $t \sim {2 Mpc \over v} \sim 7 \cdot 10^8$ 
year.  Since $\ln ({b_{\rm max} \over b_{\rm min}}) \sim 1$, we 
find $\Delta \theta \sim 6 \times 10^{-2}$, indicating that 
the flows and caustics associated with the first throughfall 
($n=1$) are only partially diffused by gravitational scattering.  
It appears unlikely that the caustics associated with flows that 
have passed several times ($n \sim$ few) through the central parts 
of the cluster survive.  On the other hand, we expect discrete 
flows and associated caustics for $n=1$, and possibly for $n= 2,3$.

Mahdavi et al. have already presented evidence for caustics 
in galaxy clusters.  When plotted as a function of position 
on the sky, the velocity dispersion and surface density of 
the galaxies in the group around NGC 5846 has a sharp drop 
at 840 kpc = $R_1$ from the group's center, consistent with 
the occurrence of the first outer caustic at that radius \cite{Mahd}.  
Furthermore, the observations of Mahdavi et al. allow a test of the
self-similar model prediction for the relationship between velocity 
dispersion and the radius $R_1$ of the first outer caustic. The model 
depends on a parameter $\epsilon$ \cite{FG} which is related to the 
slope of the power spectrum of density perturbations \cite{Dor}.  
$\epsilon$ is a slowly increasing function of the object size.  
On the scale of galaxies, $\epsilon$ is in the range 0.25 to 0.30, 
whereas for clusters it is in the range 0.35 to 0.45 \cite{STW}.  
Because NGC 5846 is a small cluster, $\epsilon$ is near 0.35.  
For $\epsilon$ = 0.3, 0.35 and 0.40, the model predicts 
respectively ${\sqrt{<v^2>} \over R_1}$ = 620, 590 and 
580 km/s~Mpc, where $\sqrt{<v^2>}$ is the velocity dispersion 
measured {\it at the center} of the cluster.  The result 
$\sqrt{<v^2>}$ = 425 km/s of Mahdavi et al. \cite{Mahd} for 
NGC 5846 agrees with the self-similar model at the 15\% level.

Since the flow of infalling dark matter forms the first 
inner caustic before it forms the first outer caustic, the 
evidence of ref. \cite{Mahd} for the first outer caustic 
of the NGC 5846 group is also evidence for its first inner 
caustic.  The former cannot exist without the latter.

We mentioned that the catastrophe structure 
of the inner caustics depends on the angular momentum 
distribution of the infalling dark matter.  If that
distribution is dominated by net overall rotation, the 
inner caustics are rings whose cross-section is an 
elliptic umbilic catastrophe.  Evidence has been 
found for caustic rings of dark matter in galaxies 
\cite{crdm,Kinn,milk,mon}, implying that the dark 
matter accreting onto galactic halos has net overall 
rotation.  Assuming that this is so, there is no reason 
to expect differently on the only somewhat larger length 
scale of galactic clusters.  Thus we are led to expect
that clusters have one or more causic rings of dark 
matter.

\section{Gravitational lensing by caustics in galaxy clusters}

Weak lensing responds to the column density distribution
$\Sigma (x,y)$ along lines of sight in the direction 
$\hat{z}$.  Caustics have precisely defined density 
profiles and hence precisely defined column density 
profiles.  Thus one may be able to answer unambiguously 
whether a feature seen in a column density map is due to 
a caustic or not.

For an outer caustic, the density is
\begin{equation}
d(r) = {A \over \sqrt{R-r}}~\Theta (R-r)
\label{outden}
\end{equation}
where $r$ is the radial coordinate, $R$ is the radius of the 
caustic, $A$ is a constant called the fold coefficient, and 
$\Theta$ is the step function.  The column density is 
\cite{Hoga,Char,Gava}
\begin{equation}
\Sigma (x) = \pi A \sqrt{2R}~\Theta (R-x)~~~\ ,
\label{outcol}
\end{equation}
for $x \ll R$.  $x$ is the projected radial coordinate.  For 
the cluster Cl 0024+1654, appropriate values for the self-similar 
infall model input parameters are: age $t$ = 9.4 Gyr, velocity 
dispersion $\sqrt{<v^2>}$ = 1300 km/s and $\epsilon = 0.4$.  The 
model predicts then 
\begin{eqnarray}
\{R_n: n = 1,2, ... 6\} = ~~~~~~~~~~~~~~~~~~~~~~~~~~~~~~\nonumber\\
(1.6,~0.9,~0.7,~0.55,~0.45,~0.40)~{\rm Mpc}
\label{outrad}
\end{eqnarray}
and
\begin{eqnarray}
\{\Sigma_n \equiv \pi A_n \sqrt{2 R_n}: n =1,2 ... 6\} 
= ~~~~~~~~~~~~\nonumber\\
(12,~11,~10,~10,~9.5,~9.5)~{M_\odot \over {\rm pc}^2}~~\ .
\label{outsig}
\end{eqnarray}
Let us briefly consider the proposal that the feature seen by Jee et al.
is due to an outer caustic in Cl 0024 + 1654.  An advantage of this
interpretation is that the circular appearance of the feature does 
not require an accident.  Since outer caustics are approximately
spherically symmetric, they appear approximately axially symmetric 
from any vantage point. However, there are serious difficulties with 
this hypothesis.  First, for the radius $R_n$ to fit the observed 
radius of the feature (400 kpc), $n$ should be of order 6; see 
Eq.~(\ref{outrad}).  But, as we discussed above, the caustics with 
such high $n$ are almost entirely degraded by gravitational scattering 
off the galaxies in the cluster.  Second, the feature identified by 
Jee et al. has, at its maximum, a column density of order 
58 ${M_\odot \over {\rm pc}^2}$ which is a 
factor five or so larger than the column density contrasts $\Sigma_n$ 
produced by outer caustics; see Eq.~(\ref{outsig}).  Finally, the
predicted profile does not fit the observed profile.  As a function 
of projected distance $x$ to the cluster center, the observed $\Sigma(x)$
shows a {\it bump} at $x \simeq$ 400 kpc, whereas the $\Sigma(x)$ due to 
outer caustics is a step function.

Caustic rings of dark matter were described in detail in 
ref.~\cite{sing}.  The properties of an axially symmetric 
caustic ring are determined entirely in terms of six parameters: 
the caustic ring radius $a$, the sizes $p$ and $q$ of its 
cross-section in the directions parallel and perpendicular 
to the plane of the ring, the speed $v$ and the mass infall 
rate per unit solid angle ${d M \over d\Omega dt}$ of the 
particles constituting the caustic, and a parameter $s$ which 
is of order $a$.  Let $z$ be the coordinate perpendicular to 
the plane of the ring and $x$ the radial coordinate at a particular 
location on the ring.  $\hat{y}$ is in the direction tangent 
to the ring.  The density $d(x,z)$ has a unique expression in 
terms of the stated parameters \cite{sing}.  It is straightforward 
to integrate $d(x,z)$ along any line in the $xz$ plane.  For the 
lines of sight perpendicular to the plane of the ring, the result 
is 
\begin{equation}
\Sigma(x) = \int dz~d(x,z)
= {4 \over v}~\sqrt{a \over s}~ 
{d M \over d\Omega dt}~{1 \over x}I\left({2 (x-a) \over s}\right)
\label{incol}
\end{equation}
where
\begin{equation}
I(\Delta) = \int_0^{\pi \over 2} d\alpha~\Theta(\alpha^2 + \Delta)
{\cos \alpha \over \sqrt{\alpha^2 + \Delta}}~~~\ .
\label{Del}
\end{equation}
$I(\Delta)$ has a logarithmic singularity
\begin{equation}
I(\Delta) \simeq {1 \over 2} \ln \left({3.2 \over |\Delta|}\right)
\label{log}
\end{equation}
when  $\Delta \rightarrow 0$.  The finite velocity dispersion
of the particles in the flow forming the caustic smoothes the 
caustic and therefore also the logarithmic singularity of 
Eq.~(\ref{log}).  Fig. 1 shows the function $I(\Delta)$, 
as well as $I(\Delta)$ averaged over two different scales.
The column density does not depend sharply on the angle of 
view.  One can show in particular that from any vantage point 
in the $xz$ plane there is a line of sight for which the column 
density is logarithmically divergent in the cold flow limit.

\section{Comparison with observation}

The self-similar model predicts $v$ and ${d M \over d\Omega dt}$
in terms of the age and velocity dispersion of the object.  For 
Cl 0024+1654, using the previously mentioned input parameters, 
this implies
\begin{equation}
\Sigma(x) = 31~{M_\odot \over {\rm pc}^2} 
~\sqrt{a \over s}~{a \over x}~I\left({2 (x-a) \over s}\right)~~~\ .
\label{clsig}
\end{equation}
To determine ${a \over s}$, information must be provided about 
the dependence of specific angular momentum on declination $\alpha$ 
near $\alpha = 0$ \cite{sing}.  In the absence of such information, 
it is only possible to state that ${a \over s}$ is of order one,
i.e. that it is likely between 0.5 and 2.  

As was mentioned, the logarithmic singularity in the column density
profile at $x = a$ is smoothed by the velocity dispersion of the dark 
matter flow which makes the caustic.  We now discuss two sources of
velocity dispersion: the presence of small scale structure in the 
dark matter falling onto the cluster, and gravitational scattering 
by inhomogeneities in the cluster.

\begin{figure}
\includegraphics[width=1.0\columnwidth]{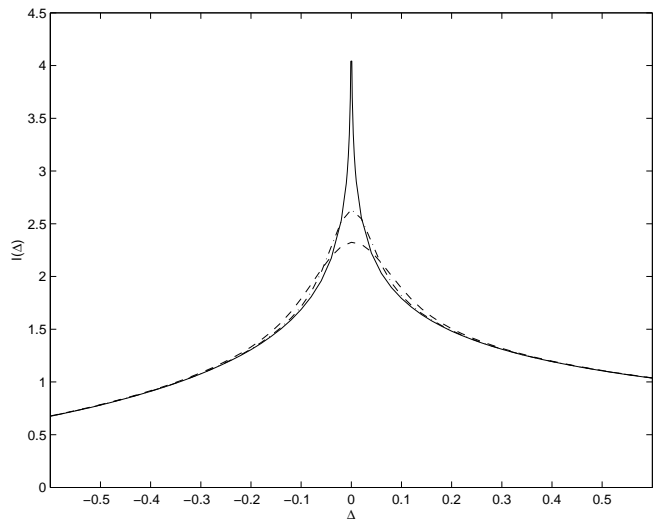}
\caption{The function $I(\Delta)$ defined by Eq.~(\ref{Del}) 
(solid line), the function $I(\Delta)$ convoluted with a 
Gaussian of FWHM 0.07 (dot-dashed), and FWHM 0.133 (dashed).}
\label{sig}
\end{figure}

In the hierarchical clustering scenario of structure formation, 
the dark matter falling onto a galaxy cluster has previously 
formed smaller scale objects.  The infalling flow of dark matter 
has effective velocity dispersion equal to the velocity dispersion 
of the small scale objects in the flow.  If this effective
velocity dispersion is $\delta v$, the spread in specific
angular momentum is $\delta \ell \sim \delta v~R$ where $R$
is the turnaround radius.  Hence the spread in the caustic 
ring radius is $\delta a \sim {\delta \ell \over v}
\sim {\delta v \over v} R$ \cite{crdm}.  For Cl 0024+1654, 
$R \sim$ 6.7 Mpc.  Since $v \sim 2 \sqrt{<v^2>} \sim$ 2600 km/s, 
the requirement that the caustic ring not be spread in radius 
over more than the observed width ($\sim$ 150 kpc) of the ring 
implies $\delta v \lesssim$ 60 km/s.  This places an upper limit 
on the size of the small scale structures composing the ring.
They cannot be larger than small galaxies.  Since no small scale 
structures are observed in the ring, $\delta v$ may be much less 
than 60 km/s.

The flow of dark matter falling onto a cluster for the first 
time is diffused through gravitational scattering at least 
somewhat by the time it forms its first inner caustic.  Using
Eq.~(\ref{scat}), we estimate that the resulting velocity 
dispersion in Cl 0024+1654 is of order $\delta v \sim 0.035 v$.
The associated spread in specific angular momentum is of order
$\delta \ell \sim a~\delta v$.  Hence the first inner caustic 
is smoothed over a minimum distance scale of order 0.035 (400 kpc) 
= 14 kpc.  This corresponds to a smoothing scale of 0.07 in 
$\Delta = {2(x-a) \over s}$ if $s=a$.  Fig.1 shows $I(\Delta)$ 
defined by Eq.~(\ref{Del}), and $I(\Delta)$ averaged over smoothing 
scales of 0.07 and 0.133 in $\Delta$, by convoluting with Gaussians 
with those full widths at half maximum (FWHM).  

The observed dark matter ring is also smoothed by the resolution 
of the Cl 0024+1654 mass reconstruction, which is limited by the 
number density of background galaxies and the finite grid size.  
If the actual ring column density were a delta function in the 
radial coordinate, it would be observed as a Gaussian bump with 
FWHM of order 27 kpc \cite{priv}.  Let us call $I_{\rm av}(\Delta)$ 
the function $I(\Delta)$ smoothed over the corresponding scale 
(0.133) in $\Delta$.  $I_{\rm av}(\Delta)$ is shown by the 
dashed curve in Fig.1 for the case $s = a$.   

The dark matter ring in Cl 0024+1654 appears as a bump in 
the graph of the azimuthally averaged column density plotted 
as a function of distance $x$ to the center of the ring \cite{Jee}.  
Fig.2 shows the data in the region of the bump.  Each data point 
is the measured average column density at the corresponding radius 
with the 1 $\sigma$ error bar \cite{Jee}.  The step in $x$ between 
successive data points is the bin size (21 kpc).  The significance 
of the bump over an assumed constant background is approximately 
8 $\sigma$ \cite{Jee}. 

\begin{figure}
\includegraphics[width=1.0\columnwidth]{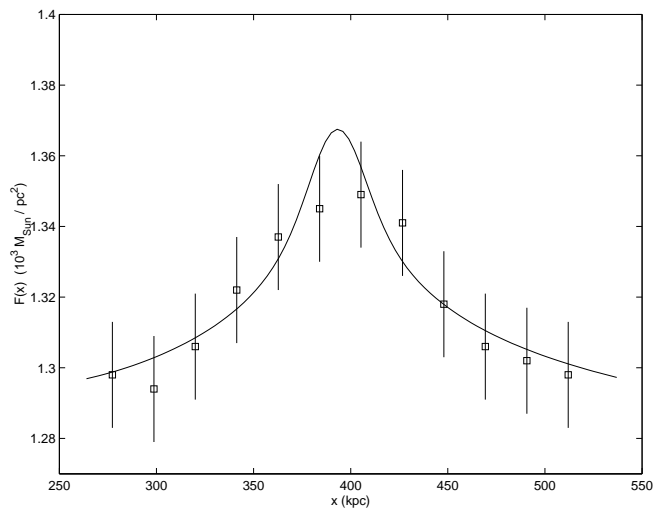}
\caption{The azimuthally averaged column density of 
Cl 0024+1654 in the range $270 < x < 520$ kpc, fitted 
to the function defined in Eq.~(\ref{fitF}).} 
\label{fit}
\end{figure}

The data points in Fig.~2 were fitted with the function 
\begin{equation}
F(x) = B~{1 \over x} + C + A~\sqrt{a \over s} 
~{a \over x} I_{\rm av}\left({2(x - a) \over s}\right)~~~\ ,
\label{fitF}
\end{equation}
where $A = 31 {M_\odot \over {\rm pc}^2}$, 
$I_{\rm av}\left({2(x - a) \over s}\right)$ is 
$I\left({2(x - a) \over s}\right)$ averaged over 
27 kpc in $x$, and $B$, $C$, $a$ and $s$ are fitting 
parameters. The first two terms describe the background 
column density due to the cluster as a whole and due to 
other matter that is smoothly distributed on the scale 
of the ring thickness.  The last term represents the 
contribution of the caustic ring of dark matter smoothed 
over the resolution of the observations.  The least squares 
fit returned $a$ = 394 kpc, $s$ = 171 kpc, 
$B = 6.2\times 10^6 {M_\odot \over {\rm pc}}$, and 
$C = 1.26~10^3~{M_\odot \over {\rm pc}^2} $.  The $\chi^2$ 
value was 2.2 for 8 degrees of freedom, indicating an 
acceptable fit.  The $\chi^2$ value is low because the data 
points do not fluctuate relative to one another as much as 
{\it independent} data points with the stated error bars 
would.  The fitted function is plotted in Fig.2.  The value 
1.5 of $\sqrt{a \over s}$, although slightly high, is acceptable 
in the context of the model since ${a \over s}$ is expected to 
be of order one.  We also fitted the data with $F(x)$ defined 
in Eq.~(\ref{fitF}) imposing the constraint $s = a$, and using
$A$, $B$, $C$ and $a$ as fitting parameters.  In this case, 
the least squares fit returned $a$ = 395 kpc, 
$A$ = 46 ${M_\odot \over {\rm pc}^2}$, 
$B = - 6.6 \times 10^6 {M_\odot \over {\rm pc}}$, and 
$C = 1.28~10^3~{M_\odot \over {\rm pc}^2} $, with $\chi^2$ = 
2.0 for 8 degrees of freedom.  The fitted function is almost 
the same as the one plotted in Fig. 2.  Thus, if one imposes 
$s = a$ as a constraint, the measured column density is 50\% 
higher than the column density predicted by the self-similar 
infall model.

The caustic ring radius $a$ is determined in the self-similar
infall model in terms of a free parameter called $j_{\rm max}$ 
which is the specific angular momentum of the particles in the 
plane of the ring.  For $a$ to equal the observed radius (395 kpc) 
of the dark matter ring, one must require $j_{\rm max} \simeq 0.45$.

\section{Conclusions}

We conclude that the dark matter ring observed by Jee et al.
has properties consistent with a caustic ring of dark matter. 
The column density agrees both in shape and overall amplitude.
At present the data are too imprecise to infer with confidence 
the nature of the observed ring.  However, our proposal makes 
a distinct prediction for the column density profile across the 
ring, as illustrated in Fig.1. This may be tested by future 
observations.  In this regard, let us emphasize that the 
${1 \over x} I_a({2(x-a) \over s})$ profile applies to each 
azimuth.  If the caustic ring interpretation is confirmed, 
an important corollary is that dark matter falls onto 
Cl 0024+1654 with net overall rotation.

~~
We thank Igor Tkachev for making available his numerical codes 
to solve the equations of the self-similar infall model.  We 
thank Leanne Duffy for useful discussions. This work was 
supported in part by the U.S. Department of Energy under 
contract DE-FG02-97ER41029.  P.S. gratefully acknowledges 
the hospitality of the Aspen Center of Physics while working 
on this project.


\begin{thebibliography}{}

\bibitem{Jee}
M.J. Jee et al., Ap. J. 661 (2007) 728.

\bibitem{bul}
D. Clowe, A. Gonzalez and M. Markevitch, Ap. J. 604 (2004) 596; 
D. Clowe et al., Ap. J. 648 (2006) L109.

\bibitem{ZuH}
J.A. ZuHone, D.Q. Lamb and P.M. Ricker, arXiv:0809.3252.

\bibitem{Ips}
P. Sikivie and J. Ipser, Phys. Lett. B291 (1992) 288.

\bibitem{crdm}
P. Sikivie, Phys. Lett. B432 (1998) 139.

\bibitem{sing}
P. Sikivie, Phys. Rev. D60 (1999) 063501.

\bibitem{inner}
A. Natarajan and P. Sikivie, Phys. Rev. D73 (2006) 023510.

\bibitem{rob}
A. Natarajan and P. Sikivie, Phys. Rev. D72 (2005) 083513.

\bibitem{FG}
J.A. Fillmore and P. Goldreich, Ap. J. 281 (1984) 1.

\bibitem{B}
E. Bertschinger, Ap. J. Suppl. 58 (1985) 39.

\bibitem{STW}
P. Sikivie, I. Tkachev and Y. Wang, Phys. Rev. Lett. 75 (1995) 2911
and Phys. Rev. D56 (1997) 1863.

\bibitem{milk}
P.\ Sikivie, Phys. Lett. B567 (2003) 1.

\bibitem{mon}
A. Natarajan and P. Sikivie, Phys. Rev. D76 (2007) 023505.

\bibitem{Kinn}
W. Kinney and P. Sikivie, Phys. Rev. D61 (2000) 087305.

\bibitem{Stiff}
D. Stiff and L. Widrow, Phys. Rev. Lett. 90 (2003) 211301.

\bibitem{simca}
A.G. Doroshkevich et al., MNRAS 192 (1980) 321;
A.A. Klypin and S.F. Shandarin, MNRAS 204 (1983) 891;
J.M. Centrella and A.L. Melott, Nature 305 (1983) 196;
A.L. Melott and S.F. Shandarin, Nature 346 (1990) 633.

\bibitem{White}
S.D.M. White and M. Vogelsberger, MNRAS 392 (2009) 281.

\bibitem{Czos}
O. Czoske et al., Astron. and Astroph. 372 (2001) 391, 
and references therein.

\bibitem{Mahd}
A. Mahdavi, N. Trentham and R.B. Tully, Astron. J. 130
(2005) 1502;  R.B. Tully, EAS Publ. Ser. 20 (2006) 191.

\bibitem{Dor}
A.G. Doroshkevich, Astrophysics 6 (1970) 320;
P.J.E. Peebles, Ap. J. 277 (1984) 470;
Y. Hoffman and J. Shaham Ap. J. 297 (1985) 16.

\bibitem{Hoga}
C. Hogan, Ap. J. 527 (1999) 42.

\bibitem{Char}
C. Charmousis et al., Phys. Rev. D67 (2003) 103502.

\bibitem{Gava}
R. Gavazzi, R. Mohayaee and B. Fort, 
Astron. and Astroph. 454 (2006) 715.

\bibitem{priv}
J. Jee, private communication.

\end{thebibliography}
\end{document}